\documentclass[conference]{IEEEtran}
\IEEEoverridecommandlockouts
\usepackage{cite}
\usepackage{amsmath,amssymb,amsfonts}
\usepackage{algorithm}
\usepackage{graphicx}
\usepackage{textcomp}
\usepackage{multirow}
\usepackage[hyphens,spaces,obeyspaces]{url} 
\usepackage{hyperref}
\usepackage{amsmath}
\usepackage{float}
\usepackage[noend]{algpseudocode} 
\usepackage{booktabs}
\usepackage{easyReview}
\usepackage{tabularx}
\usepackage{listings}
\usepackage{xcolor} 

\usepackage{caption}
\usepackage{subcaption}
\usepackage[export]{adjustbox}[2011/08/13]

\usepackage{etoolbox}
\makeatletter
\patchcmd{\@maketitle}
  {\addvspace{0.5\baselineskip}\egroup}
  {\addvspace{-2\baselineskip}\egroup}
  {}
  {}
\makeatother

\makeatletter
\newcommand{\linebreakand}{%
  \end{@IEEEauthorhalign}
  \hfill\mbox{}\par
  \mbox{}\hfill\begin{@IEEEauthorhalign}
}
\makeatother

\newcolumntype{L}{>{\centering\arraybackslash}m{2cm}}
\newcolumntype{S}{>{\centering\arraybackslash}m{1cm}}
\newcolumntype{Z}{>{\centering\arraybackslash}m{0.7cm}}
\newcolumntype{M}{>{\centering\arraybackslash}m{1.5cm}}

\definecolor{codegreen}{rgb}{0,0.6,0}
\definecolor{codegray}{rgb}{0.5,0.5,0.5}
\definecolor{codepurple}{rgb}{0.58,0,0.82}
\definecolor{backcolour}{rgb}{1,1,1}
\lstdefinestyle{mystyle}{
    backgroundcolor=\color{backcolour},   
    commentstyle=\color{codegreen},
    keywordstyle=\color{magenta},
    numberstyle=\tiny\color{codegray},
    stringstyle=\color{codepurple},
    basicstyle=\ttfamily\footnotesize,
    breakatwhitespace=false,         
    breaklines=true,                 
    captionpos=t,                    
    keepspaces=true,                 
    numbers=left,                    
    numbersep=5pt,                  
    showspaces=false,                
    showstringspaces=false,
    showtabs=false,                  
    tabsize=2
}
\algdef{SE}[VARIABLES]{Variables}{EndVariables}
   {\algorithmicvariables}
   {\algorithmicend\ \algorithmicvariables}
\algnewcommand{\algorithmicvariables}{\textbf{global variables}}

\def\BibTeX{{\rm B\kern-.05em{\sc i\kern-.025em b}\kern-.08em
    T\kern-.1667em\lower.7ex\hbox{E}\kern-.125emX}}
\begin{document}

\title{Machine-Learning-Driven Runtime Optimization of BLAS Level 3 on Modern Multi-Core Systems\\
\vspace{-0.5em}
}

\author{\IEEEauthorblockN{Yufan Xia}
\IEEEauthorblockA{\textit{The Chinese University of Hong Kong}\\
Hong Kong SAR, China \\
xiayufan12345@outlook.com}
\and
\IEEEauthorblockN{Giuseppe Maria Junior Barca}
\IEEEauthorblockA{\textit{The University of Melbourne}\\
Melbourne, Australia \\
giuseppe.barca@unimelb.edu.au}\\
}

\maketitle

\begin{abstract}

  BLAS Level 3 operations are essential for scientific computing, but finding the optimal number of threads for multi-threaded implementations on modern multi-core systems is challenging. We present an extension to the Architecture and Data-Structure Aware Linear Algebra (ADSALA) library that uses machine learning to optimize the runtime of all BLAS Level 3 operations. Our method predicts the best number of threads for each operation based on the matrix dimensions and the system architecture. We test our method on two HPC platforms with Intel and AMD processors, using MKL and BLIS as baseline BLAS implementations. We achieve speedups of 1.5 to 3.0 for all operations, compared to using the maximum number of threads. We also analyze the runtime patterns of different BLAS operations and explain the sources of speedup. Our work shows the effectiveness and generality of the ADSALA approach for optimizing BLAS routines on modern multi-core systems.

\end{abstract}

\section{Introduction}

The linear algebra subroutines in the Basic Linear Algebra Subprograms (BLAS) \cite{blackford2002updated} form the backbone of scientific computing.
Due to the critical role of BLAS, great effort has been devoted to improving the performance its subroutines.
The automatically tuned linear algebra software (ATLAS) and its improvements are the first batch of auto-tuning efforts on optimising linear algebra operations; they are able to auto-tune the linear algebra operation codes by searching over parameters like blocking factor, loop order, and partial storage location on each specific computer architecture \cite{whaley1998automatically,gunnels2001family,vuduc2001statistical}. Later in the 2000s, the self-optimised linear algebra routine (SOLAR) attempted the analytical modelling of the timing of multi-process linear algebra operations \cite{gunnels2001systematic,valsalam2002framework,demmel2005self,herrero2006framework,cuenca2002towards,Cuenca2004}.

Within BLAS, Level 3 (L3) operations, which are concerned with matrix-matrix linear algebra, are the most complex and computationally demanding \cite{blackford2002updated}.  In addition to the General Matrix Multiplication (GEMM), BLAS L3 includes the Symmetric Rank-k Update (SYR), Symmetric Rank-2k Update (SYR2), Triangular Matrix-Matrix Multiplication (TRMM), and Triangular Rank-k Update (TRSM) subroutines. 

While there has been significant work in optimizing single-threaded CPU performance of BLAS L3 by fine-tuning the size of matrix blocks for various system architectures, similar efforts for optimizing these routines specifically for multi-core CPU have been less frequent due to the high complexity and large variety of the underlying computer architectures. 

Work has been carried out by using established optimization techniques such as parameter tuning and blocking\cite{Katagiri2017, catalan2019case}. The Parallel Linear Algebra for Scalable Multi-Core Architectures (PLASMA) \cite{camara2013empirical} library uses polynomial regression to optimise the number of threads or block size used based on empirical features, obtaining performance comparable to the Intel$^{\circledR}$ MKL for Cholesky decomposition. Peise \emph{et al.} adopted a polynomial regression to model the dense linear algebra run times, and they applied additional techniques to boost the performance of this simple model \cite{peise2012performance}. 

However, choosing the number of threads that minimises the execution time of a given BLAS remains challenging and largely unsolved due to the underlying diversity and complexity of modern shared-memory computer architectures. 

To tackle this challenge, recent research conducted by Xia \emph{et al.} \cite{ADSALA}, which was integrated in the open-source Architecture and Data-Structure Aware Linear Algebra library, has employed a systematic machine learning (ML) methodology to significantly reduce the runtime of Single-precision General Matrix Multiplication (SGEMM) operations. This approach leverages an ML model to dynamically select the optimal number of threads for minimizing the execution time of GEMM for a specific input configuration and computer system architecture. The ML model itself undergoes training during the installation phase, and this training process is customized to the particular computer system architecture in use. Subsequently, during runtime, the ML model is employed to make predictions regarding the most efficient number of threads for a given task. An important advantage of this approach is its adoption of existing GEMM implementations, treating them as black boxes. This work has inspired a novel auto-tuning method in Graph Neural Network training \cite{lin2024argo}.

{In this study, we have expanded the runtime optimization capabilities of the ADSALA library to encompass all single- and double-precision BLAS L3 operations, and we present the performance results achieved. Furthermore, we have incorporated an automatic model selection feature during the installation process for each distinct BLAS subroutine. This addition aids in identifying and using the most appropriate machine learning algorithm for each subroutine on every machine where the library is installed.}
We test our method on two HPC platforms with Intel and AMD processors, using multi-threaded MKL and BLIS as baseline BLAS implementations. Our software implementation enables us to speed up all BLAS L3 subroutines, notwithstanding the runtime overhead of ML evaluations. 

On the Gadi supercomputer located at the National Computational Infrastructure (NCI) and on the Setonix supercomputer located at the Pawsey Supercomputing Centre (please refer to section \ref{sec: Experimentation Information} for details about NCI and Pawsey), we achieve speedups of 1.5 $\times$ to 3.0 $\times$ compared to using the maximum number of threads with hyperthreading enabled or disabled. 
For comparison, the most recent method relevant to our study employs pure polynomial regression to model and determine the number of threads for the PLASMA QR routine\cite{camara2013empirical}. {This method reported average speed improvements of 1\%, 13\%, and 27\% across three different platforms. The sizes of the matrices used in these tests ranged from 2000 to 7000, which aligns closely with the range we tested in our study.}

We also analyze the runtime patterns of different ADSALA BLAS L3 runs and discuss the sources of speedup.  The ADSALA library is provided as open-source implementation for the community to use and extend.

The remainder of this article is structured as follows. We provide  background information concerning BLAS L3 operations and ML in Section \ref{sec: Background}. We review the improved software design in Section \ref{sec: Software Workflow}, and discuss the adopted ML methods in Section \ref{sec: Machine Learning Methods}. Section \ref{sec: Experimentation Information} details the experimentation platform and settings. We present and analyze the performance of the ML models and software speedup in Section \ref{sec: Performance Analysis}. Section \ref{sec: Conclusion} concludes.

\begin{table*}[!ht]
  \centering
  \caption{Specifications of BLAS level III subroutines.}
  \label{tab:BLAS III}

{
\scriptsize
\sffamily

\begin{tabular}{SSSMSMSM}

\toprule
\\[-2ex]
& {\textbf{}} & {\textbf{Matrix A}} &  {\textbf{}} & {\textbf{Matrix B}} & {\textbf{}} & {\textbf{Matrix C}} & {\textbf{}}\\

\cmidrule(r){3-7}
& dims & shape & type & shape & type & shape & type\\
\midrule

\textbf{GEMM} &  3 & m $\times$ k & regular & k $\times$ n & regular & m $\times$ n & regular\\

\textbf{SYMM} &  2 & m $\times$ m & symmetric & m $\times$ n & regular & m $\times$ n & regular\\

\textbf{SYRK} & 2 & n $\times$ k & regular & n $\times$ k & regular & n $\times$ n & symmetric\\

\textbf{SYR2K} &  2 & n $\times$ k & regular & n $\times$ k & regular & n $\times$ n & symmetric\\

\textbf{TRMM} & 2 & m $\times$ m & triangular & m $\times$ n & regular & \textemdash & \textemdash \\

\textbf{TRSM} & 2 & m $\times$ m & triangular & m $\times$ n & regular & \textemdash & \textemdash \\

\bottomrule
\end{tabular}
}



  \vspace{-1em}
\end{table*}

\begin{table}[!ht]
  \centering
  \caption{Comparisons of ML model characteristics.}
  \label{tab:model_comp}
  
{
\scriptsize
\sffamily

\begin{tabular}{SLSMM}

\toprule
\\[-2ex]
{\textbf{Model Catagories}} & {\textbf{Models}} &  {\textbf{Parametric}} & {\textbf{Good with Data Imbalance}} & {\textbf{Data Size Requirement}} \\
\midrule
\multirow{ 5}{0.15\columnwidth}{\textbf{Linear Models}} & Linear Regression & \multirow{ 5}{0.15\columnwidth}{Yes} & \multirow{ 5}{0.15\columnwidth}{No} & \multirow{ 1}{0.15\columnwidth}{Medium} \\
\cmidrule(r){2-2}\cmidrule(r){5-5}
& ElasticNet &  &  & \multirow{ 1}{0.15\columnwidth}{Medium} \\
\cmidrule(r){2-2}\cmidrule(r){5-5}
& Bayesian Regression &  &  & \multirow{ 1}{0.15\columnwidth}{Small} \\
\midrule
\multirow{ 8}{0.15\columnwidth}{\textbf{Tree \\Based Models}} & Decision Tree & \multirow{ 8}{0.15\columnwidth}{No} & \multirow{ 8}{0.15\columnwidth}{Yes} & \multirow{ 8}{0.15\columnwidth}{Medium}\\
\cmidrule(r){2-2}
& XGBoost &  &  &  \\
\cmidrule(r){2-2}
& AdaBoost &  &  &  \\
\cmidrule(r){2-2}
& Random Forest &  &  &  \\
\cmidrule(r){2-2}
& LightGBM &  &  &  \\
\midrule
\multirow{ 3}{0.15\columnwidth}{\textbf{Other Models}} & SVM Regressor & \multirow{ 3}{0.15\columnwidth}{No} & \multirow{ 3}{0.15\columnwidth}{No} & \multirow{ 1}{0.15\columnwidth}{Small}\\
\cmidrule(r){2-2}\cmidrule(r){5-5}
& KNN Regressor &  & &\multirow{ 1}{0.15\columnwidth}{Medium}\\

\bottomrule
\end{tabular}
}

  \vspace{-1em} 
\end{table}

\section{Background}
\label{sec: Background}
In this Section, we introduce BLAS L3 subroutines, ML algorithms, and related data preprocessing techniques.

\subsection{BLAS Level III Subroutines}

BLAS L3 includes six matrix-matrix operations, including GEMM, SYMM, SYR, SYR2, TRMM, and TRSM \cite{blackford2002updated}. These operations have different number of matrix operands, matrix shapes, and some matrix operands are required to be symmetric or triangular. These requirements are listed in Table \ref{tab:BLAS III}. These subroutines provide a unified interfaces for different BLAS implementations, thus enables users to switch between different BLAS implementations without changing their code.

The performance of BLAS L3 subroutines is highly dependent on the underlying implementation and the system architecture. For example, the GEMM routine is the most commonly used BLAS Level III subroutine, and it is usually the most optimised one for BLAS implementations. However, as reported, even GEMM has the issue where the maximum thread number, which we defined to be equal to the number of available CPU cores times the hyper-threading level, may not provide the best performance \cite{ADSALA}. While this issue has not been reported to be present in other BLAS subroutines, in later sections, we will show that the other BLAS L3 subroutines show a similar behaviour.

\subsection{Machine Learning Algorithms}

To build a high-performance ML model, we need to sift through suitable potential candidates. We provide a brief overview of the dataset we will use (for more details on data collection, please refer to subsection \ref{subsec: Data Gathering}) to help justify our selection of candidate models that best fit the data characteristics. The dataset for a given BLAS L3 operation typically comprises approximately $10^3$ data points spanning 4-15 dimensions, with most data features exhibiting a skewed distribution. Despite being a relatively small dataset with low-dimensional data, the relationship between the features and the label can be quite complex due to the expected non-linear relation stemming from the polynomial time complexity of most BLAS L3 operations and the intricacies of multi-core computer architectures. Given these attributes, we present several ML algorithms as potential candidates in Table \ref{tab:model_comp}.

Linear algorithms produce simple but high speed ML models \cite{deisenroth2020mathematics}. They usually suffer from low predictive accuracy for non-linear mappings. Due to the nature of our task, evaluation speed is an essential factor, so linear ML models may enable pragmatic BLAS L3 speedups. We include linear regression, ElasticNet and Bayesian regression as our candidates \cite{tibshirani1996regression,bishop2003bayesian}. 
 
We also include both Decision Tree and tree ensemble models as candidates. The Decision Tree algorithm is non-parametric, it uses a set of rules to map each data instance to a discrete class or a continuous value \cite{bishop2006pattern}. Random Forest, AdaBoost, XGBoost and LightGBM are more complex algorithms who are built upon an ensemble of Decision Trees. These tree ensemble algorithms generally can reduce bias and variances compared to a single Decision Tree and thus can produce a more promising result but with higher evaluation time.

Support Vector Machine (SVN) and k-Nearest-Neighbors (kNN) are non-parametric algorithms but are designed to work well for high dimensions and unknown data structures, respectively \cite{awad2015support}. Since we have low data dimensionality, they might not demonstrate their advantages when learning from our dataset; kNN is also known to be slow in evaluation \cite{bhatia2010survey}. We will evaluate their performance for completeness.

\begin{figure*}[!htbp]
  \centering 
  \begin{subfigure}{.55\textwidth}
    \centering
    \includegraphics[width=0.9\linewidth]{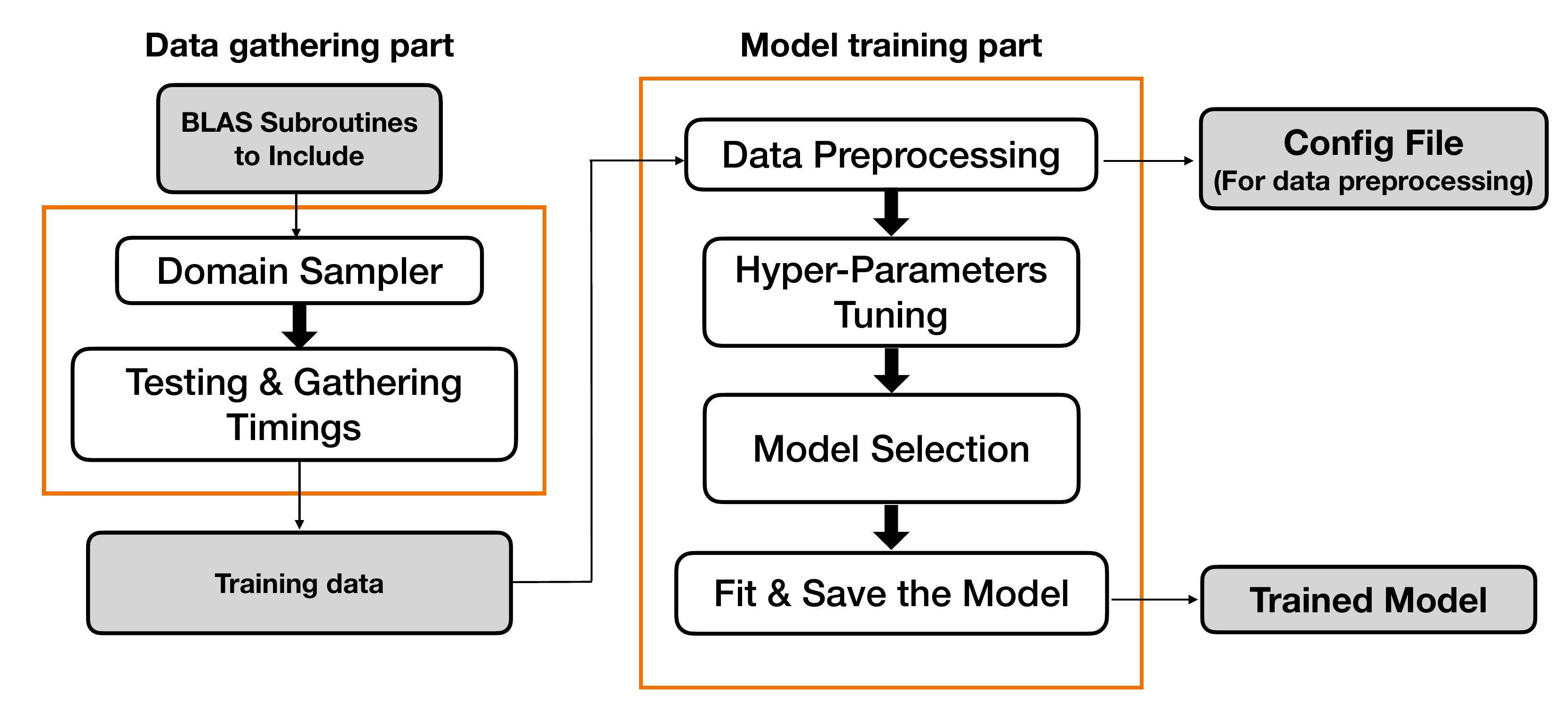}
    \caption{The installation workflow of ADSALA GEMM. Upon ADSALA installation, the library performs two sub-parts shown in the diagram. In the end, two files containing the configurations together with the production-ready ML model will be saved for later use at runtime.}
    \label{fig:Lib Structure}
  \end{subfigure}%
  \hspace{1cm} 
  \begin{subfigure}{.35\textwidth}
    \centering
    \includegraphics[width=0.9\linewidth]{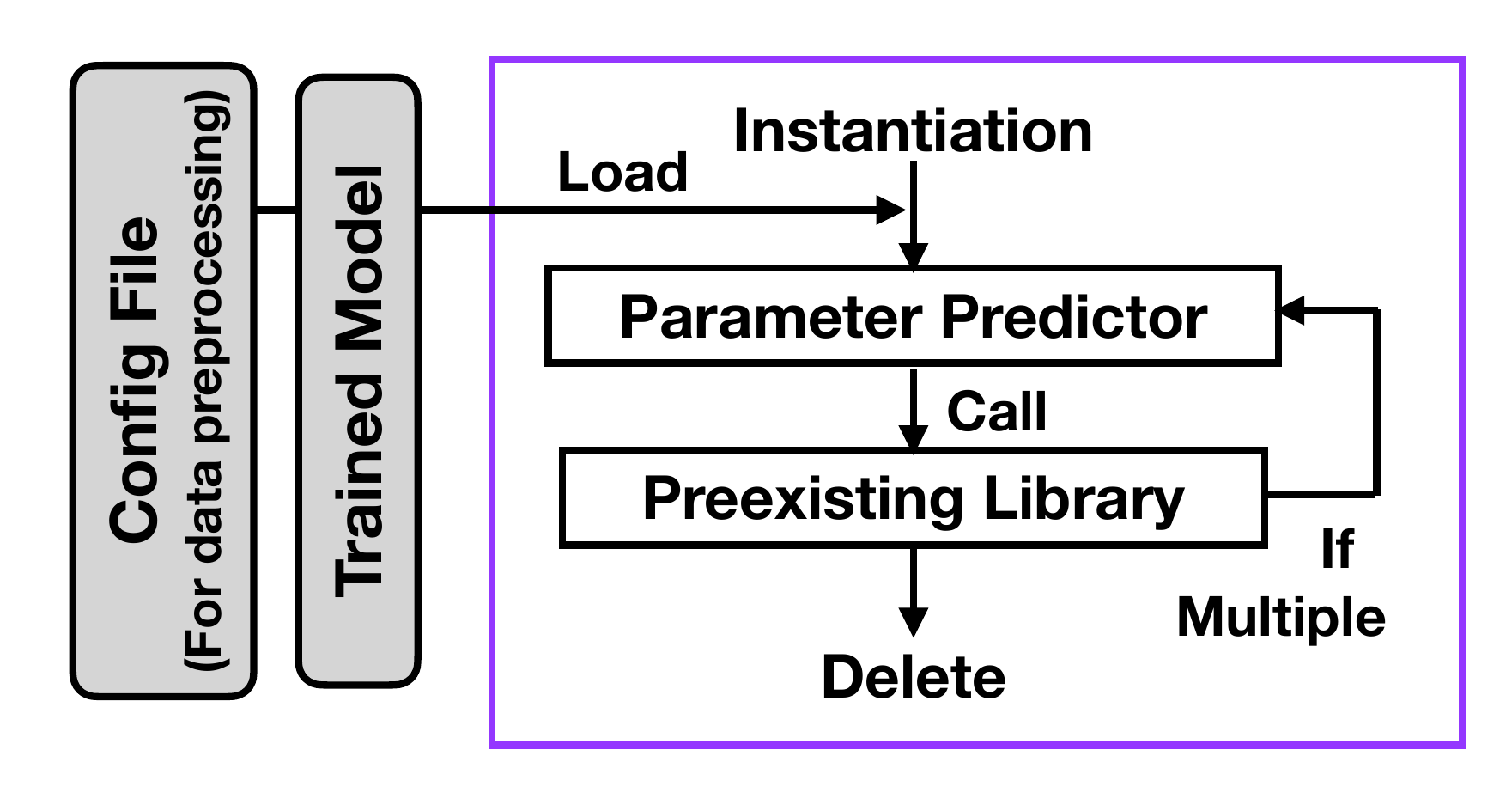}
    \caption{The runtime workflow of ADSALA for BLAS level III subroutines. Configuration files and trained ML models outputted during installation (see Fig. \ref{fig:Lib Structure}) are used by this runtime library.}
    \label{fig: Predictor Design}
  \end{subfigure}
  \caption{The software design of ADSALA.}
  \label{fig: Software Design}
\end{figure*}

\subsection{Data Preprocessing Techniques}
\label{subsec: Data Preprocessing Techniques}

Data outliers can affect the predictive performance of ML models. While statistical methods can effectively remove global outliers, they often fail to detect local outliers. the Local Outlier Factor (LOF), a density-based method, overcomes this limitation by assigning each data point a degree of isolation score based on its surrounding points \cite{han2022data,10.1145/335191.335388}. We use LOF to identify and remove these outliers from our dataset.

{The Yeo-Johnson transformation is a data transformation technique that remaps each feature value such that the feature distribution is near-Gaussian, thus improving the predictive performance of ML algorithms that assume normality, such as linear regression, ElasticNet, and Bayesian regression.} Unlike the original Box-Cox transformation, which requires all feature values to be positive, Yeo-Johnson accepts non-positive values and provides stable parameter estimation \cite{sakia1992box,weisberg2001yeo}. Yeo-Johnson transformation uses a parameter $\lambda$ to control the strength of transformation. We apply Yeo-Johnson with maximum likelihood estimation (MLE) for parameter estimation for the data transformation, thereby automating the ML workflow.


\section{Software Workflow}
\label{sec: Software Workflow}

Our software, upon installation, collects runtime data for each of BLAS L3 subroutine and different combinations of thread numbers, input matrices dimensions, and BLAS packages availability. This data trains an ML model to predict the optimal thread number for given dimensions of the GEMM input. The final product is an ADSALA library that dynamically selects the optimal thread number at runtime, adapting to different HPC platforms, BLAS packages, and subroutines.

Figure \ref{fig:Lib Structure} and \ref{fig: Predictor Design} illustrate our software design procedures. The software comprises data gathering, model training, and a runtime library. The first two parts are executed at installation time, while the third part is typically used at runtime when the user program links to our library.

\subsection{Installation Workflow}
During installation, the software gathers training data through experimentation. It quasi-randomly (\emph{vide infra}) samples from the domain of the specific subroutine matrices dimensions, passes them to a timing program that runs and times the corresponding BLAS operations. The timings are stored as training data for the ML model. The timing data is then preprocessed and utilised to tune the hyper-parameters of the ML model.

\subsection{Runtime Workflow}
\label{subsec: Runtime Workflow}
Upon instantiation, the ML models and its configuration are loaded into memory. At runtime, when the user program calls a BLAS function, the ML model predicts the optimal thread number on-the-fly and runs the implementation with that thread number. The ML model and configuration are wrapped in a C++ class that can be destroyed after the last BLAS call to free memory. 

To avoid redundant ML model evaluations for two consecutive BLAS call with identical input dimensions, our software remembers the input to the last BLAS call and its correlation ML prediction. If the current BLAS input dimensions are the same as the previous, the software will simply read and apply the predictions from the responsible class attributes without re-evaluation.


\section{Machine Learning Methods}
\label{sec: Machine Learning Methods}

In this section, we build upon the original ADSALA proof-of-concept library \cite{ADSALA} and extend it to support all BLAS L3 subroutines. We first introduce the mechanism of ML predictions, then we detail the methods for data collection, data preprocessing, model selection, and model training process in our software. Since our goal is a library that works on different HPC architectures and BLAS packages, the following methodologies can be applied to any HPC architectures and BLAS packages.

\subsection{Mechanism for Predictions}
For a given BLAS subroutine and a given combination of input dimensions, the software selects the optimal number of threads by {first predicting the duration of BLAS subroutine} associated with the possible number of threads. It then selects the thread number with the shortest predicted runtime for the ensuing BLAS subroutine execution. 


\subsection{Data Gathering}
\label{subsec: Data Gathering}

{Two paragraphs that were in this section are combined to eliminate repetitions mentioned by R1}

In order to ensure that our method is effective for BLAS L3 subroutines across a variety of matrix shapes and sizes within memory constraints, including slim/square and big/small matrices, we need to sample domains that are evenly distributed across the space. To achieve this, we employ a scrambled Halton sequence to produce a quasi-random sequence with low discrepancy for data sampling \cite{MascagniChi+2004+435+442}. The upper bound of the sum size of matrices to 500 MB\footnote{Some BLAS subroutines have their output matrix overwrite the input matrix (TRMM and TRSM). We calculate the sum memory size as the sum of input/output matrices, ignoring the overwritten matrices.}.  
 Given that the samples can have two or three domains, we opt for the scrambled Halton sequence over the regular Halton sequence to reduce the correlation between dimensions \cite{MascagniChi+2004+435+442}. We utilize bases 2, 3, and 4 for generating the sequence for dimensions $m$, $k$, and $n$ and bases 2, 3 for dimensions $m$, $n$ (or $n$, $k$). These samples are then input into the timing program to gather their runtime data.

\subsection{Feature Engineering and Data Preprocessing}
\label{subsec: Data Preprocessing} 

Table \ref{tab:additional features} shows the features used for the ML model. We created two set of features for subroutines with three matrix size parameter and two matrix size parameter with respect to the computational complexity, their memory footprint, and the multi-thread speedup. 
 
The execution time for BLAS subroutines is a function of the matrix dimensions and the number of threads ($nt$), and it varies depending on the architecture. For three-dimension subroutines, the terms $m*k$, $k*n$, $m*n$, and $m*k+k*n+m*n$ correspond to the sizes of matrices $A$, $B$, $C$, and the total memory size in single-/double-precision words, respectively. Similar terms are constructed for two-dimension subroutines as shown in the second group in Table \ref{tab:additional features}. These memory terms are directly related to the number of memory operations and hence, influence the runtime of various BLAS subroutines. Specifically, these terms are dominant in the serial runtime for smaller matrix dimensions. In three-dimension subroutines, the cubic term $m*k*n$ is proportional to the number of floating-point operations performed and tends to dominate the runtime in serial execution for larger matrix dimensions. In parallel execution, the FLOP workloads are distributed across threads, resulting in terms like ${m*n*k}/nt$.

For feature transformation, outlier removal, feature selection, and hyper-parameter tuning stages, we use the same methodologies as in the original ADSALA library {that helps improving the speed and also predictive performance of the model}\cite{ADSALA}. {We employ the Yeo-Johnson transformation to approximate the distribution of features to a near-Gaussian distribution, which is beneficial for model learning.}\footnote{{In our tests, we observed an enhancement in the performance of these models when the Yeo-Johnson Transformation was applied, resulting in a 10-20\% decrease in the Root Mean Square Error (RMSE) for Linear Regression. The application of it does not affect the performance of other candidate models much.}} This transformation identifies the most suitable parameter values for $\lambda$ and adjusts the transformation’s impact on each feature using the MLE method \cite{weisberg2001yeo}. Following this transformation, we carry out a standardisation process on features to ensure they all operate on a similar scale \cite{geron2019hands}. {Subsequently, we eliminate features that have correlation coefficients with other features exceeding a threshold of $80\%$ to eliminate the potentially redundant features among the candidates.} For each correlation feature pair, we remove the feature with the larger total correlation with the other features. Following this, the hyper-parameter tuning is performed for all models to compare model performance.

\begin{table}[t!]
\vspace{-1.5em}
  \centering
  \caption{List of available features for BLAS subroutines with two or three matrix dimensions. nt stands for number of threads.}
  \label{tab:additional features} 
  
{
    \scriptsize
    \sffamily
    
    \begin{tabular}{p{0.03\textwidth}cccc}
    \toprule
    &   {\multirow{2}{0.20\columnwidth}{\textbf{Three Dimensions $m$, $k$, $n$}}}     && \multirow{ 2}{0.20\columnwidth}{\textbf{Two Dimensions $m$, $n$ }} & \\ 
    \\[2ex] 

    \midrule
    1 & m &    ${\text{m}}/{\text{nt}}$ & m             \\[3pt]
    2 & k&    ${\text{k}}/{\text{nt}}$  & n            \\[3pt]
    3 & n&    ${\text{n}}/{\text{nt}}$   & nt          \\[3pt]
    4 & nt&    ${\text{m*k}}/{\text{nt}}$ &  m*n              \\[3pt]
    5 & m*k&    ${\text{m*n}}/{\text{nt}}$ &  memory\_footprint            \\[3pt]
    6 & m*n&    ${\text{k*n}}/{\text{nt}}$  & m/nt            \\[3pt]
    7 & k*n&    ${\text{m*k*n}}/{\text{nt}}$ &  n/nt             \\[3pt]
    8 & m*k*n&    memory\_footprint / ${\text{nt}}$ & m*n/nt            \\[3pt]
    9 & memory\_footprint& &        memory\_footprint / nt \\
    \bottomrule    
    \end{tabular}
}


  \vspace{-1em}
\end{table}


\subsection{Model Selection}
\label{subsec: Model Tuning and Selection}

{Given that our method’s ultimate goal is to achieve optimal runtime speedup, the model selection process incorporates both the predictive performance of the model and the speed of model evaluation with a potential trade-off between these two.} This approach is akin to the proof-of-concept work conducted with ADSALA \cite{ADSALA} where the speedup is estimated using the formula: 
$$s = \frac{t_{\text{original}}}{t_{\text{ADSALA}}+t_{\text{eval}}}$$
In this context, $t_{\text{ADSALA}}$ is the runtime of the BLAS subroutine with predicted threads, $t_{\text{eval}}$ is the ML model evaluation time, and $t_{\text{original}}$ is the runtime with maximum threads. The evaluation time $t_{\text{eval}}$ is measured by averaging multiple runs. The ML model with the highest average estimated speedup $s$ across all BLAS subroutines is selected.


\section{Experimentation Information}
\label{sec: Experimentation Information}

This Section presents the supercomputing platforms used for experimentation and the setup of the experiments on those platforms.

\subsection{Experimentation Platforms}
\label{subsec: Experimentation Platforms}

We conducted our experiments on two supercomputing platforms.

\subsubsection{Setonix} 

\begin{figure}[h!]
  \vspace{-1em}
  \centering
  {\includegraphics[width=0.85\columnwidth]{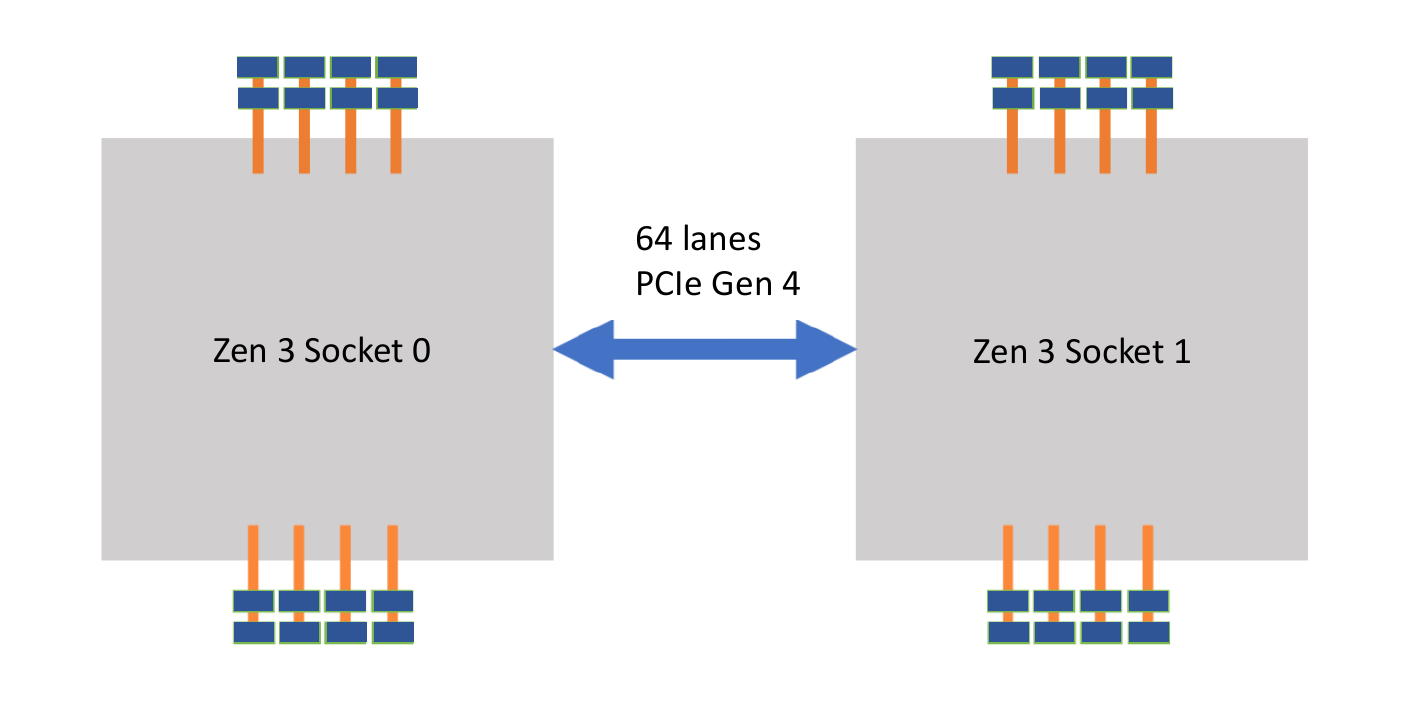}}
  \caption{A schematic diagram for the 2-socket EPYC CPU configuration on Setonix.}
  \label{fig: Setonix Socket}
  \vspace{-1em}  

\end{figure}

Setonix is a supercomputer housed at the Pawsey Supercomputing Research Centre in Australia. Each of its compute nodes has specific features, as depicted in Fig. \ref{fig: Setonix Socket}.

\begin{itemize}
  \item Each compute node is equipped with two CPU sockets, each housing an AMD$^{\circledR}$ EPYC 64-Core \textit{Milan} CPU (2.55 GHz). The CPUs support hyper-threading, enabling up to 256 threads to run concurrently per compute node. Each \textit{Milan} CPU comprises eight modules, with each module containing eight Zen 3 cores and a dedicated 32 MB level three cache that is shared among these cores.
  \item The system has a memory capacity of 256GB, organized into eight NUMA domains, with four NUMA domains per socket. Each socket supports eight memory channels.
\end{itemize}

\subsubsection{Gadi}

\begin{figure}[h!]
  \centering
  {\includegraphics[width=0.9\columnwidth]{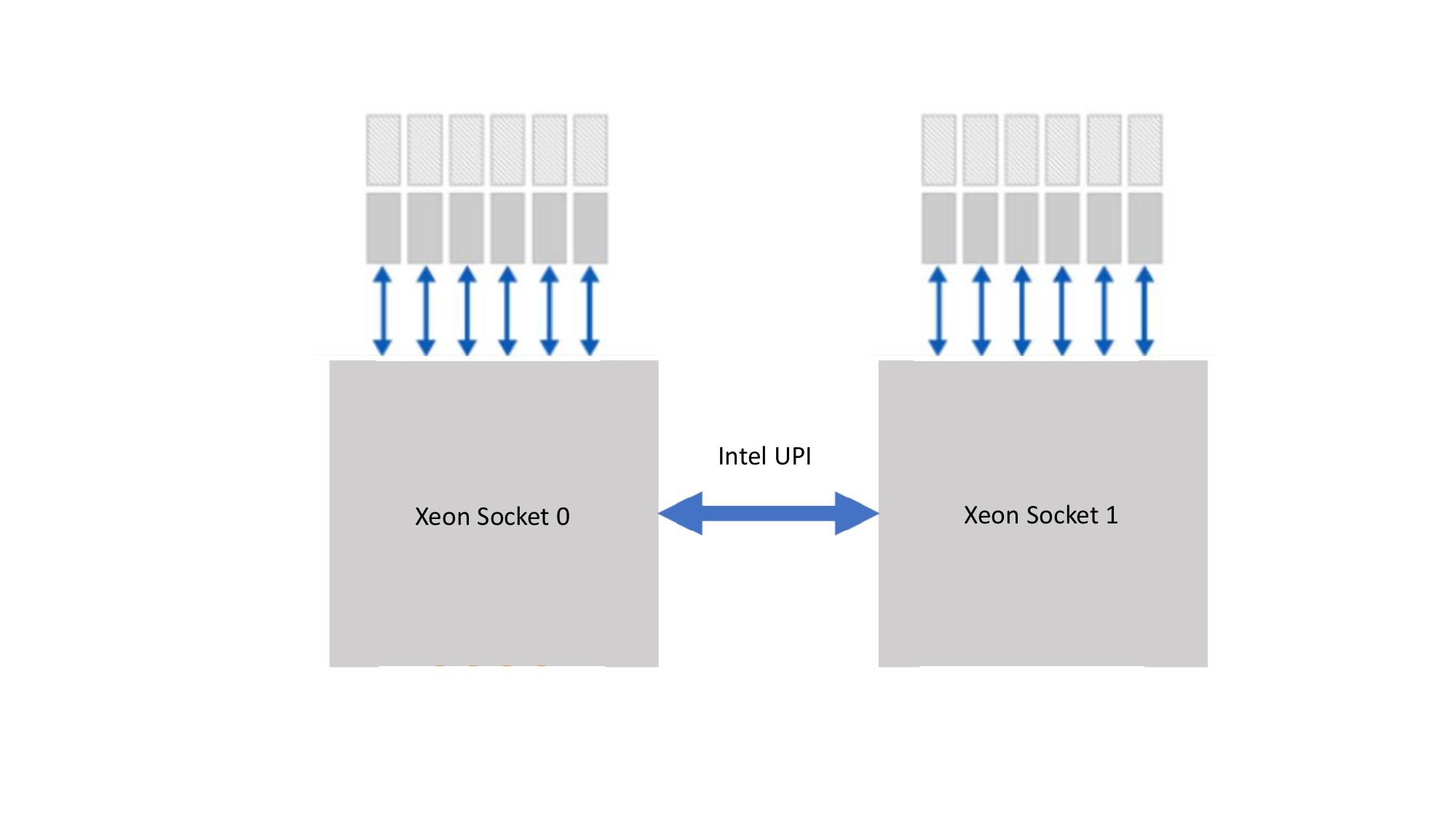}}
  \caption{A schematic diagram of the 2-socket Cascade Lake CPU configuration; sockets are connected using Intel$^{\circledR}$ UPI (Ultra Path Interconnect). }
  \label{fig: Gadi Socket}
  \vspace{-1em}
\end{figure}

Gadi is a supercomputer located at the Australian National Computational Infrastructure. Figure \ref{fig: Gadi Socket} illustrates the features of each of Gadi's compute nodes:
\begin{itemize}
  \item Each compute node has two CPU sockets, each containing an Intel$^{\circledR}$ Xeon 24-Core \textit{Cascade Lake} CPU (Platinum 8274, 3.2 GHz). The CPUs support hyper-threading, enabling up to 96 threads to run concurrently per compute node.
  \item The system has a memory capacity of 192GB, organized into four NUMA domains, with two NUMA domains per socket. Each socket supports six memory channels.
  \end{itemize}

\subsection{Experimentation Setup}
For the Setonix supercomputer, which uses AMD$^{\circledR}$ processors, we will use the performance of BLIS\footnote{ \url{https://developer.amd.com/amd-aocl/blas-library/ }} as the baseline for measuring performance improvements. For the Gadi supercomputer, which uses Intel$^{\circledR}$ processors, we will use the performance of MKL\footnote{ \url{https://www.intel.com/content/www/us/en/develop/documentation/get-started-with-mkl-for-dpcpp/top.html} } as the baseline for measuring performance improvements. For convenience in comparion of performance with the proof-of-concept ADSALA work, we will use the same experimental settings which can be found in \cite{ADSALA}.

\begin{table*}[!htbp]
  \centering
  \begin{minipage}{.4\textwidth}
    \caption{Model performance and estimated speedups for ML models on Setonix.}
    \label{tab:estimated_speedup_setonix}
    \scriptsize
\sffamily

\begin{tabular}{ll}
    \toprule
    \textbf{subroutine} &       \textbf{best\_model} \\
    \midrule
         dgemm & LinearRegression \\
         dsymm &     XGBRegressor \\
        dsyr2k &     XGBRegressor \\
         dsyrk &     XGBRegressor \\
         dtrmm &     XGBRegressor \\
         dtrsm &     XGBRegressor \\
         \midrule
        sgemm &     XGBRegressor \\
         ssymm &     XGBRegressor \\
        ssyr2k &     XGBRegressor \\
         ssyrk &     XGBRegressor \\
         strmm &     XGBRegressor \\
         strsm &     XGBRegressor \\
    \bottomrule
\end{tabular}

  \end{minipage}%
  \hspace{1cm}
  \begin{minipage}{.4\textwidth}
    \caption{Model performance and estimated speedups for ML models on Gadi.}
    \label{tab:estimated_speedup_gadi}
    \scriptsize
\sffamily

\begin{tabular}{ll}
    \toprule
    \textbf{subroutine} &       \textbf{best\_model} \\
    \midrule
         dgemm &         BayesianRidge \\
         dsymm & RandomForestRegressor \\
        dsyr2k &          XGBRegressor \\
         dsyrk &      LinearRegression \\
         dtrmm &          XGBRegressor \\
         dtrsm &      LinearRegression \\
         \midrule
         sgemm &          XGBRegressor \\
         ssymm & RandomForestRegressor \\
        ssyr2k &          XGBRegressor \\
         ssyrk &          XGBRegressor \\
         strmm &          XGBRegressor \\
         strsm &      LinearRegression \\
    \bottomrule
    \end{tabular}
    
  \end{minipage}
\end{table*}


\begin{table*}[!htbp]
  \vspace{-1em}
  \caption{{Detailed statistics for model performance and estimated speedups for ML models on Gadi. The bold values of each column represents the best entry in that column.}}
   \centering
  \label{tab:detailed stats speedup}
  \scriptsize
\sffamily

\begin{tabular}{crLLLLLL}
    \toprule
    \multicolumn{8}{c}{\textbf{dgemm Gadi}} \\
    \midrule
    {} &  \textbf{Model} &    \textbf{Normalised Test RMSE} &  \textbf{Ideal mean speedup} &  \textbf{Ideal aggregate speedup}  &  \textbf{Model evaluation time in $\mu$s} &  \textbf{Estimated mean speedup} &  \textbf{Estimated aggregate speedup} \\
    \midrule
    &  \textbf{Linear Regression} &       0.93 &	{1.13} &	1.01 &	15.3 &	1.12 &	1.01 \\
    &  \textbf{ElasticNet} &     1.00 &                1.07 &                   0.94  &           10.63 &                    1.07 &                       0.94 \\
    &  \textbf{Bayes Regression} &       0.93 &                \textbf{1.13} &                   \textbf{1.01}  &            8.11 &                   \textbf{1.13} &                       \textbf{1.01} \\
    &  \textbf{Decision Tree} &       0.32 &                0.82 &                   0.36  &            \textbf{8.02} &                   0.82 &                   0.36 \\
    &  \textbf{Random Forest} &       0.24 &                1.08 &                   0.97  &        983.09 &                    1.01 &                       0.94 \\
    &  \textbf{AdaBoost} &       0.52 &                0.84 &                   0.46  &          89.77 &                    0.84 &                       0.46 \\
    &  \textbf{KNN} &   0.28 & 1.07  & 0.95   &  6449.32   &   0.78 &  0.78\\
    &  \textbf{XGBoost} &       \textbf{0.23} &                1.1 &                   0.98  &          290.76 &                    1.08 &                       0.97 \\
\end{tabular}

\begin{tabular}{crLLLLLL}
    \toprule
    \multicolumn{8}{c}{\textbf{dsymm Gadi}} \\
    \midrule
    {} &  \textbf{Model} &    \textbf{Normalised Test RMSE} &  \textbf{Ideal mean speedup} &  \textbf{Ideal aggregate speedup}  &  \textbf{Model evaluation time in $\mu$s} &  \textbf{Estimated mean speedup} &  \textbf{Estimated aggregate speedup} \\
    \midrule
    &  \textbf{Linear Regression} &       1.00&	1.49&	1.86&	6.49&	1.49	&1.86 \\
    &  \textbf{ElasticNet} &     1.00 &                1.45 &                   1.8  &           4.78 &                    1.45 &                       1.8 \\
    &  \textbf{Bayes Regression} &       1.00 &                1.49 &                   1.86  &            4.65 &                    1.49 &                       1.86 \\
    &  \textbf{Decision Tree} &       0.46 &                1.07 &                   0.54  &            \textbf{4.56} &                   1.07 &                   0.54 \\
    &  \textbf{Random Forest} &       \textbf{0.15} &                1.72 &                   1.89  &        563.59 &                    \textbf{1.62} &                       \textbf{1.86} \\
    &  \textbf{AdaBoost} &       0.61 &                0.76 &                   0.36  &          69.88 &                    0.75 &                       0.36 \\
    &  \textbf{KNN} &   0.18 & \textbf{1.76}  & \textbf{1.95}   &  2775.13   &   1.42 &  1.79\\
    &  \textbf{XGBoost} &       0.18 &                1.65 &                   1.82  &          1046.46 &                    1.5 &                       1.77 \\
\end{tabular}

\begin{tabular}{crLLLLLL}
    \toprule
    \multicolumn{8}{c}{\textbf{ssyrk Gadi}} \\
    \midrule
    {} &  \textbf{Model} &    \textbf{Normalised Test RMSE} &  \textbf{Ideal mean speedup} &  \textbf{Ideal aggregate speedup}  &  \textbf{Model evaluation time in $\mu$s} &  \textbf{Estimated mean speedup} &  \textbf{Estimated aggregate speedup} \\
    \midrule
    &  \textbf{Linear Regression} &       1.00&	0.95&	0.89&	6.85&	0.95	&0.89 \\
    &  \textbf{ElasticNet} &     1.00 &                0.99 &                  \textbf{0.96}  &           5.07 &                    0.99 &                       \textbf{0.96} \\
    &  \textbf{Bayes Regression} &       1.00 &                0.95 &                   0.89  &            4.97 &                    0.95 &                       0.89 \\
    &  \textbf{Decision Tree} &       0.33 &                0.56 &                   0.28  &            \textbf{4.89} &                   0.56 &                   0.28 \\
    &  \textbf{Random Forest} &       0.09 &                \textbf{1.2} &                   0.81  &        2324.21 &                    0.96 &                       0.77 \\
    &  \textbf{AdaBoost} &       0.48 &                0.53 &                   0.26  &          62.94 &                    0.52 &                       0.26 \\
    &  \textbf{KNN} &   0.11 & 1.2  & 0.85   &  1760.56   &   1 &  0.81\\
    &  \textbf{XGBoost} &       \textbf{0.09} &                1.15 &                   0.82  &          446.43 &                    \textbf{1.08} &                       0.81 \\
    
\end{tabular}

\begin{tabular}{crLLLLLL}
    \toprule
    \multicolumn{8}{c}{\textbf{strsm Gadi}} \\
    \midrule
    {} &  \textbf{Model} &    \textbf{Normalised Test RMSE} &  \textbf{Ideal mean speedup} &  \textbf{Ideal aggregate speedup}  &  \textbf{Model evaluation time in $\mu$s} &  \textbf{Estimated mean speedup} &  \textbf{Estimated aggregate speedup} \\
    \midrule
    &  \textbf{Linear Regression} &       1.00&	1.04&	0.96&	8.84&	\textbf{1.04}	&0.96 \\
    &  \textbf{ElasticNet} &     1.00 &                1.02 &                   0.95  &           6.39 &                    1.02 &                       0.95 \\
    &  \textbf{Bayes Regression} &       1.00 &                1.04 &                   0.96  &            5.72 &                    1.04 &                       0.96 \\
    &  \textbf{Decision Tree} &       0.31 &                0.64 &                   0.23  &            \textbf{4.94} &                   0.64 &                   0.23 \\
    &  \textbf{Random Forest} &       0.06 &                1.2 &                   \textbf{1.1}  &        2191.65 &                    0.96 &                       1.01 \\
    &  \textbf{AdaBoost} &       0.46 &                0.58 &                   0.27  &          119.84 &                    0.57 &                       0.27 \\
    &  \textbf{KNN} &   0.08 & 1.2  & 1.09   &  1688.15   &   1 &  \textbf{1.02}\\
    &  \textbf{XGBoost} &       \textbf{0.07} &                \textbf{1.21} &                   1.05  &          1354.14 &                    {1.04} &                       {0.99} \\
    \bottomrule
\end{tabular}
\end{table*}

\begin{figure*}[!t]
  \centering 
  \vspace{-1em}
     \begin{subfigure}[b]{1.9\columnwidth}
         \centering
        \makebox[1.0\columnwidth][c]{\includegraphics[width=1.06\columnwidth]{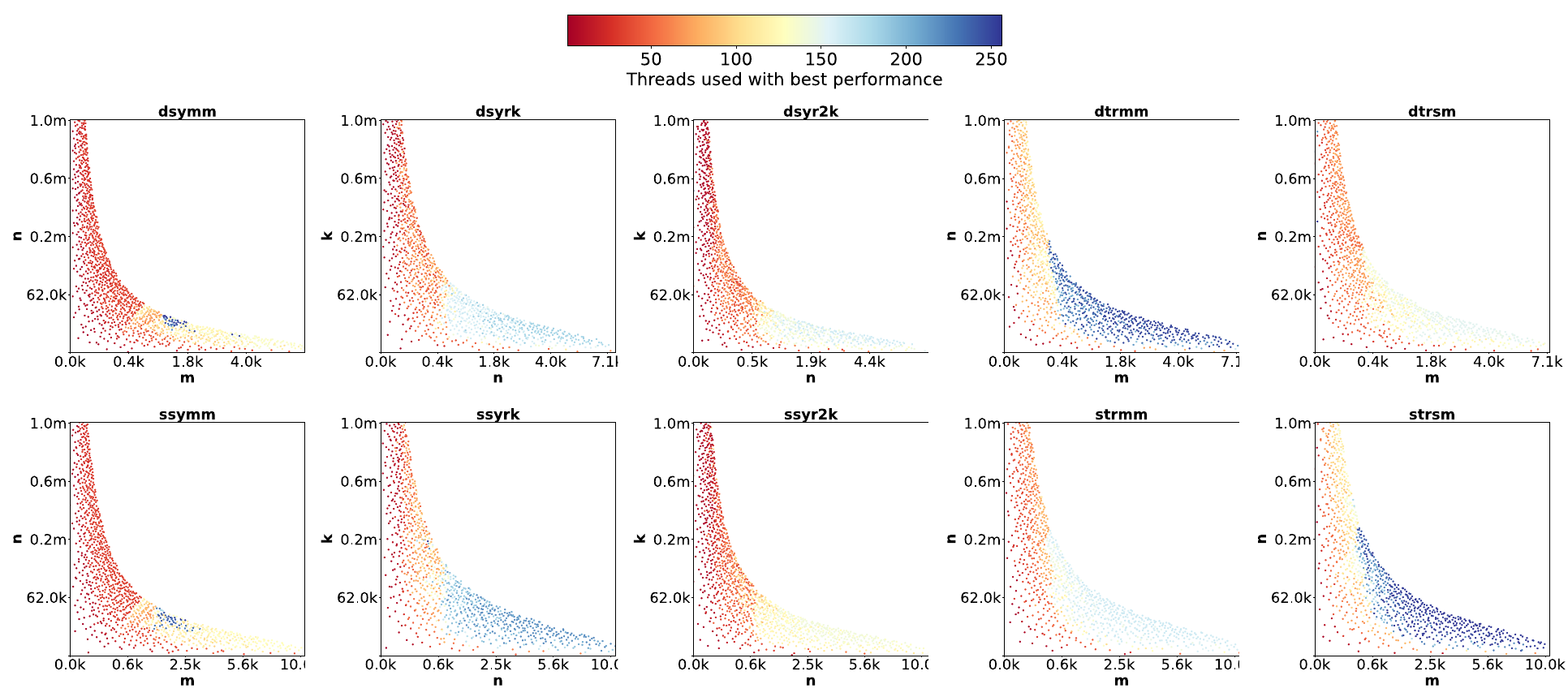}}
        \vspace{-2.2em}
        \subcaption{Setonix}
        \label{fig: data collection Setonix} 
     \end{subfigure}
    \vspace{-1.5em}
     \begin{subfigure}[b]{1.9\columnwidth}
         \centering
        \makebox[\columnwidth][c]{\includegraphics[width=1.06 \columnwidth]{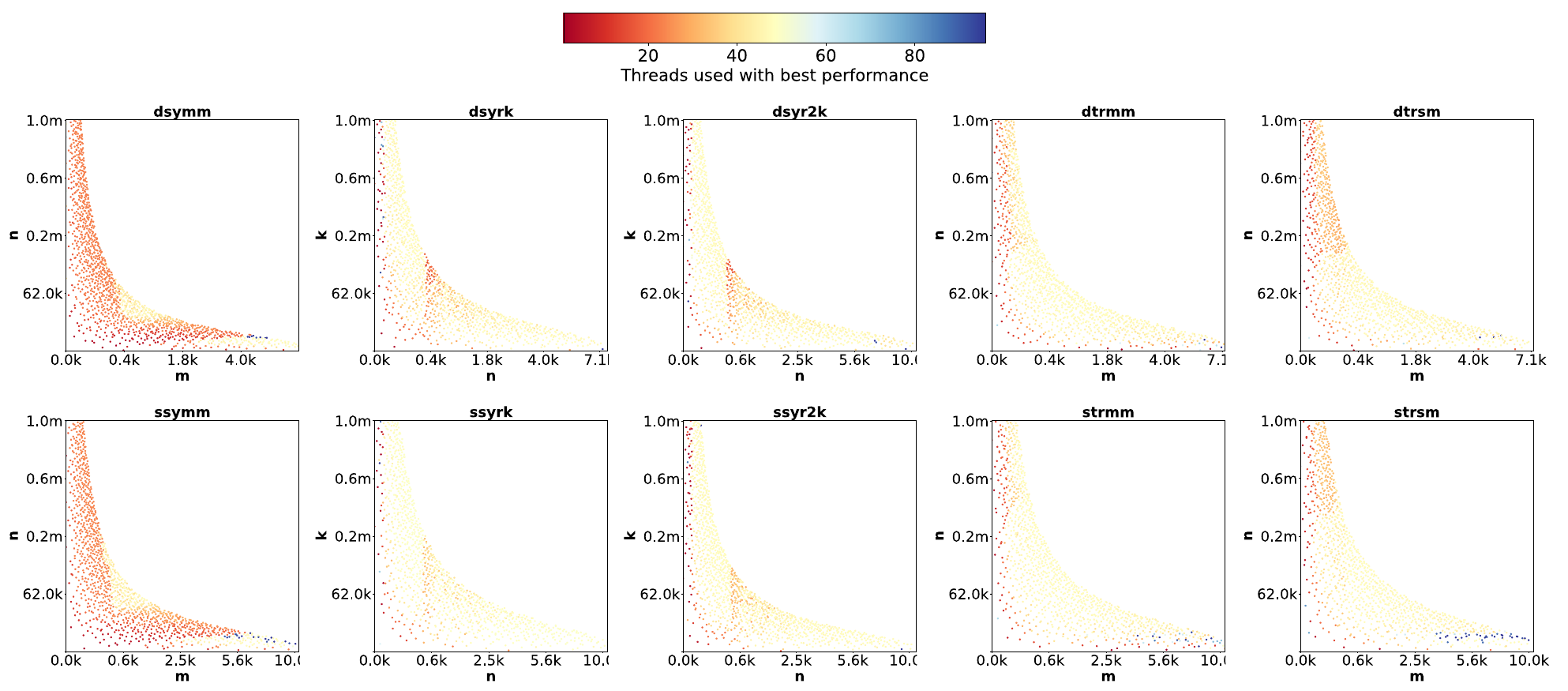}}
        \vspace{-2.2em}
        \subcaption{Gadi}
        \label{fig: data collection Gadi} 
        \vspace{1.5em}
     \end{subfigure}

  \caption{Heatmap of the optimal number of threads on Setonix and Gadi, concerning all BLAS level III subroutines except GEMM. The horizontal and vertical axes use a square root scale.} 
  \label{fig: data collection 2d} 
   \vspace{-1.8em}
\end{figure*} 
\begin{figure}[!t]
  \centering 
  \vspace{-1em}
     \begin{subfigure}[b]{1\columnwidth}
         \centering
        \makebox[\columnwidth][c]{\includegraphics[width=1.06\columnwidth]{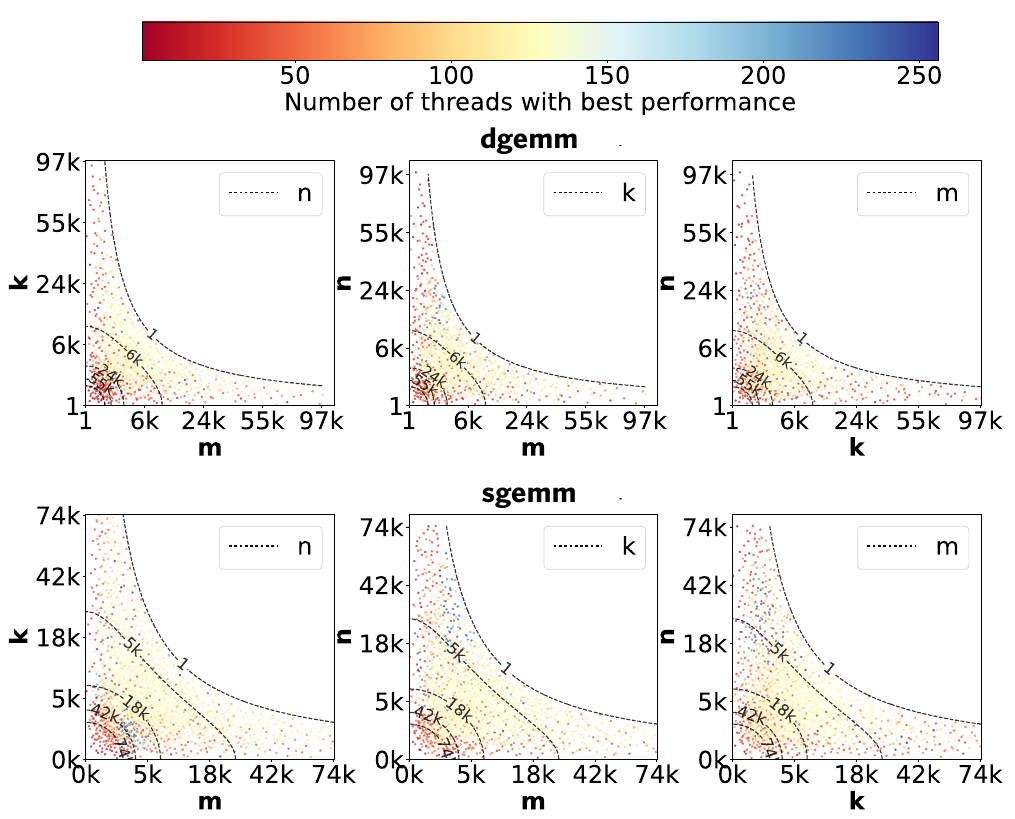}}
        \vspace{-2em}
        \subcaption{Setonix}
        \label{fig: data collection Setonix} 
     \end{subfigure}
    \vspace{-1.5em}
     \begin{subfigure}[b]{1\columnwidth}
         \centering
        \makebox[\columnwidth][c]{\includegraphics[width=1.06 \columnwidth]{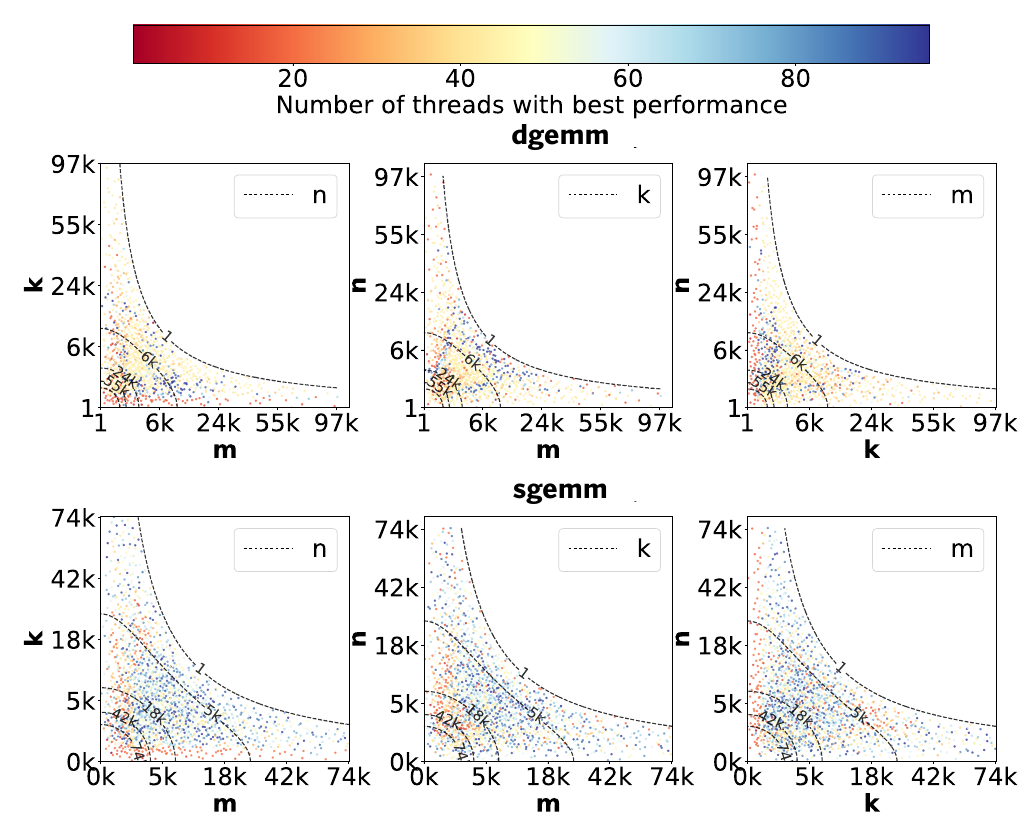}}
        \vspace{-2em}
        \subcaption{Gadi}
        \label{fig: data collection Gadi} 
        \vspace{1.5em}
     \end{subfigure}
  \caption{Heatmap of the optimal number of threads on Setonix and Gadi. The horizontal and vertical axes use a square root scale. The dashed lines on each sub-graph are contour lines of the sampling domain with each label showing the value of the third dimension.} 
  \label{fig: data collection} 
   \vspace{-1.8em}
\end{figure}

\begin{figure*}[!t]
  \centering 
  \vspace{-1em}
     \begin{subfigure}[b]{1.9\columnwidth}
         \centering 
        \makebox[\columnwidth][c]{\includegraphics[width=1.06\columnwidth]{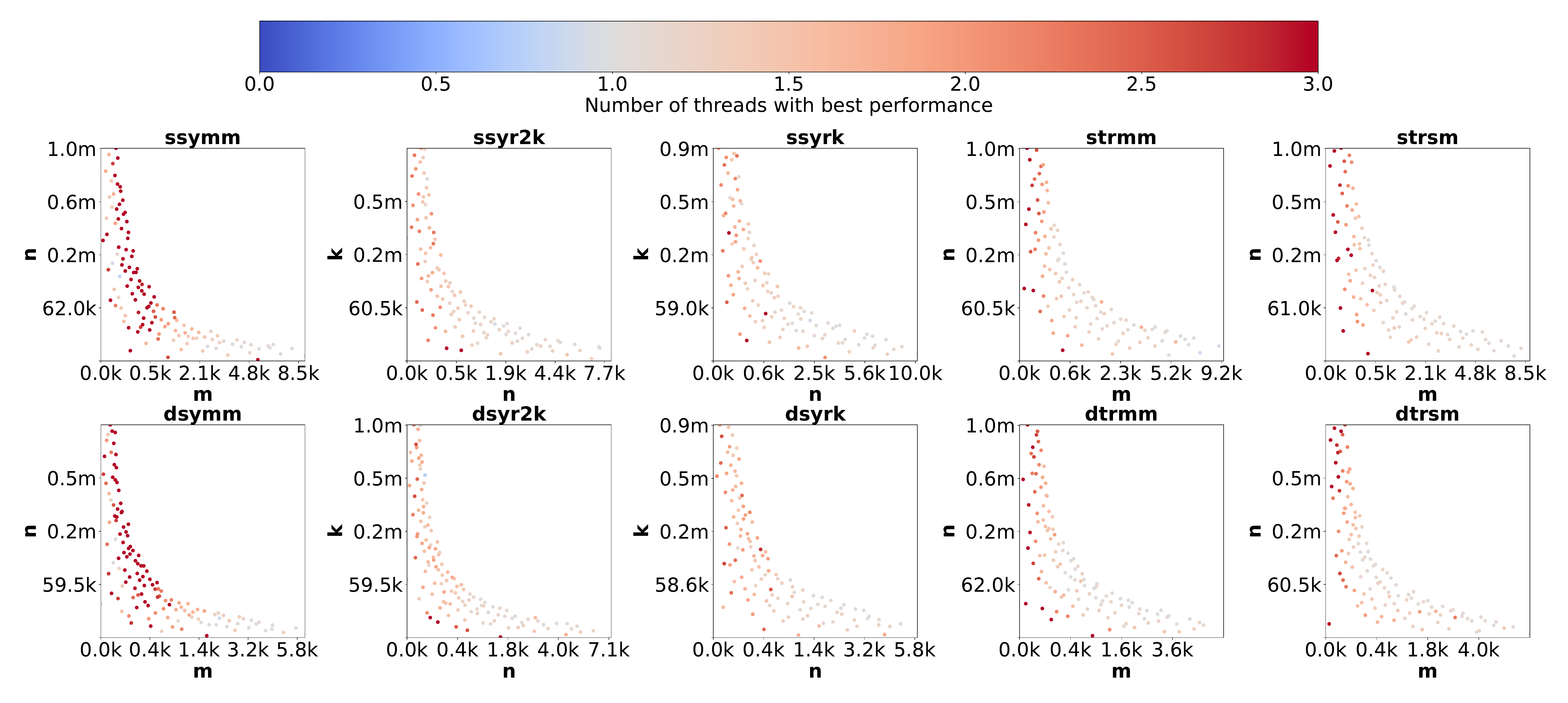}}
        \vspace{-2.2em}
        \subcaption{Setonix}
        \label{fig: 2dim speedup Setonix} 
     \end{subfigure}
    \vspace{-1.5em}
     \begin{subfigure}[b]{1.9\columnwidth}
         \centering
        \makebox[\columnwidth][c]{\includegraphics[width=1.06 \columnwidth]{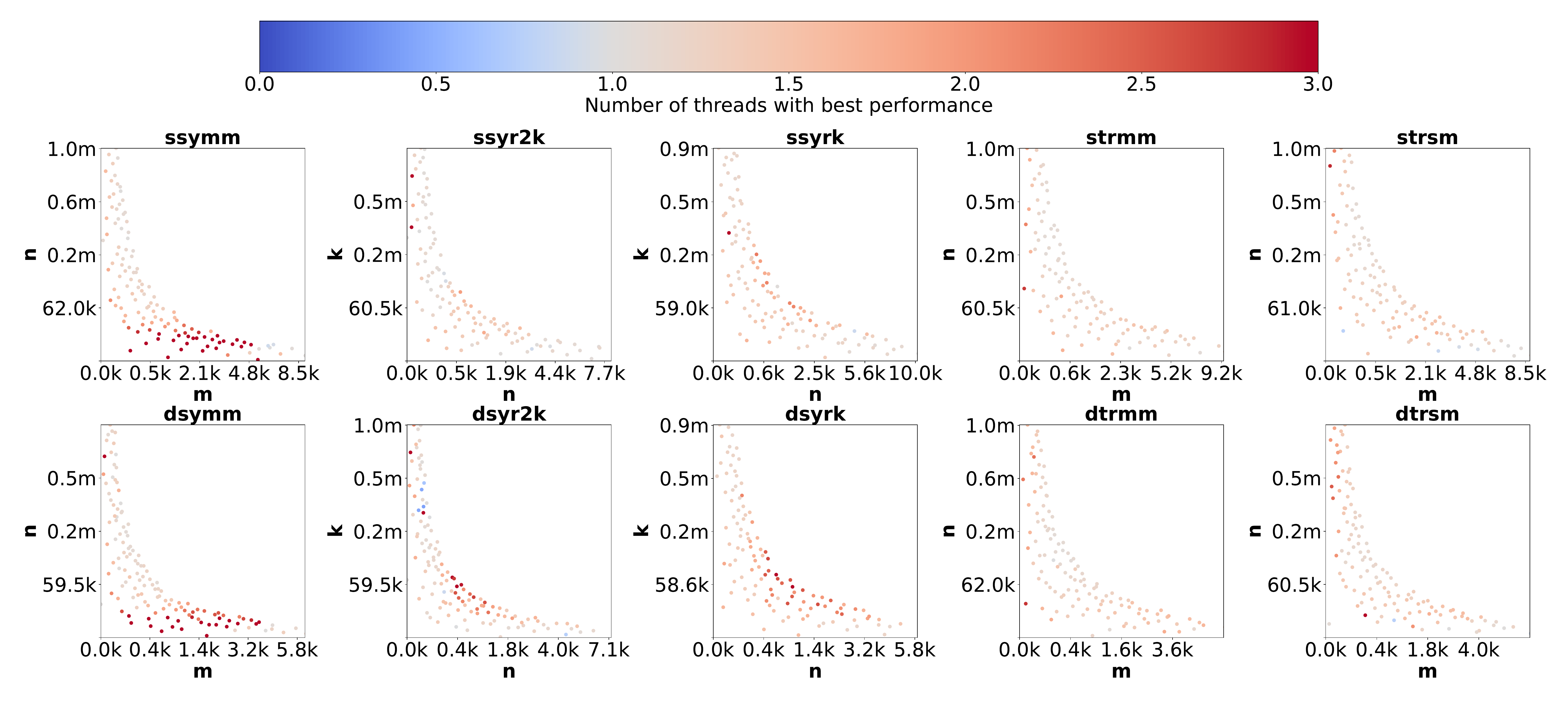}}
        \vspace{-2.2em}
        \subcaption{Gadi}
        \label{fig: 2dim speedup Gadi} 
        \vspace{1.5em}
     \end{subfigure}

  \caption{Heatmap of the testing speedup with respect to matrix sizes, concerning all BLAS level III subroutines except GEMM. The horizontal and vertical axes use a square root scale.} 
  \label{fig: 2dim speedup} 
   \vspace{-1.8em}
\end{figure*}

\begin{table}[!htbp]
  \centering
  \vspace{-0.5em}
  \caption{Speedup statistics on Setonix and Gadi with hyper-threading}
  \vspace{-1em}
  \label{tab: performance}

  \begin{subtable}{\columnwidth}
    \centering 
    \caption{Setonix}
    \label{tab: performance 500M Setonix}
    \scriptsize
\sffamily

\begin{tabular}{SZZZZZZZ}
    \toprule
     {} & \textbf{Mean} & \textbf{Std} & \textbf{Min} & \textbf{25\%} & \textbf{50\%} & \textbf{75\%} & \textbf{Max}\\
    \midrule
    {\textbf{dgemm}}   &1.54 & 0.66 & 0.87 & 1.16 & 1.29 & 1.74 &  4.79\\
    {\textbf{sgemm}}   & 1.32 & 0.41 & 0.76 & 1.05 & 1.18 & 1.37 & 9.05\\
    {\textbf{dsymm}}   & 2.89 & 1.80 & 0.82 & 1.38 & 2.36 & 4.09 & 8.46 \\
    {\textbf{ssymm}}   &2.22  &1.69  &0.34  &1.04  &1.60  &2.88  &7.42  \\
    {\textbf{dsyr2k}}  &1.48  &0.46  &0.94  &1.17  &1.36  &1.56  &3.59  \\
    {\textbf{ssyr2k}}  &1.53  &0.49  &0.75  &1.20  &1.43  &1.72  &3.79  \\
    {\textbf{dsyrk}}   &1.46  &0.46  &0.92  &1.15  &1.28  &1.69  &3.54  \\
    {\textbf{ssyrk}}   &1.55  &0.42  &0.98  &1.22  &1.46  &1.82 &2.80  \\
    {\textbf{dtrmm}}   &1.61  &0.82  &0.87  &1.13  &1.27  &1.78  &6.51  \\
    {\textbf{strmm}}   &1.67  &0.92  &0.94  &1.10  &1.34  &1.81  &7.13  \\
    {\textbf{dtrsm}}   &1.71  &1.07  &0.99  &1.13  &1.27  &1.77  &7.43  \\
    {\textbf{strsm}}   &1.68  &1.21  &0.98  &1.12  &1.34  &1.85 &12.38   \\
    \bottomrule  
\end{tabular}

  \end{subtable}
  
  \vspace{1em}
  \begin{subtable}{\columnwidth}
    \centering
    \caption{Gadi}
    \label{tab: performance 500M Gadi}
    \scriptsize
\sffamily

\begin{tabular}{SZZZZZZZ}
    \toprule
     {} & \textbf{Mean} & \textbf{Std} & \textbf{Min} & \textbf{25\%} & \textbf{50\%} & \textbf{75\%} & \textbf{Max}\\
    \midrule
    {\textbf{dgemm}}   & 1.27 & 0.55 & 0.41 & 1.01 & 1.13 & 1.29 & 4.25 \\
    {\textbf{sgemm}}   & 1.07 & 0.70 & 0.88 & 1.00 & 1.00 & 1.02 & 3.01\\
    {\textbf{dsymm}}   & 2.28 & 1.89 & 0.88 & 1.18 & 1.43 & 2.50 & 12.08 \\
    {\textbf{ssymm}}   & 2.16 & 1.98 & 0.99 & 1.18 & 1.30 & 2.22 & 11.05 \\
    {\textbf{dsyr2k}}  & 1.28 & 0.38 & 0.90 & 1.08 & 1.20 & 1.37 & 4.27 \\
    {\textbf{ssyr2k}}  & 1.47 & 0.60 & 0.41 & 1.14 & 1.26 & 1.61 & 4.52 \\
    {\textbf{dsyrk}}   & 1.40 & 0.36 & 0.86 & 1.20 & 1.30 & 1.47 & 3.98 \\
    {\textbf{ssyrk}}   & 1.65 & 0.47 & 1.11 & 1.29 & 1.45 & 1.96 &3.03 \\
    {\textbf{dtrmm}}   &1.30 &0.24 &1.03 &1.16 &1.25 &1.34 &2.75 \\
    {\textbf{strmm}}   &1.35 &0.26 &0.98 &1.17 &1.30 &1.47 &2.78 \\
    {\textbf{dtrsm}}   &1.31 &0.25 &0.75 &1.17 &1.29 &1.41 &2.86 \\
    {\textbf{strsm}}   &1.40 &0.40 &0.74 &1.19 &1.33 &1.47 &4.40 \\
    \bottomrule
\end{tabular}

  \end{subtable}

  \vspace{-1em}
\end{table}

\begin{figure}[!t]
  \centering 
  \vspace{-1em}
     \begin{subfigure}[b]{1\columnwidth}
         \centering
        \makebox[\columnwidth][c]{\includegraphics[width=1.06\columnwidth]{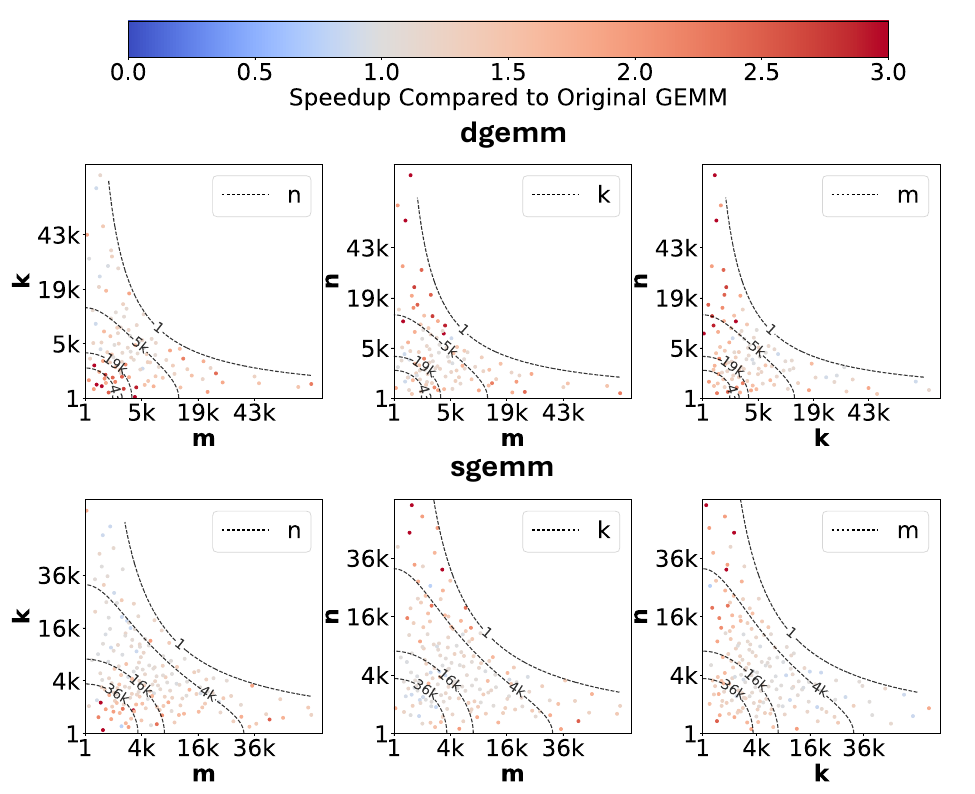}}
        \vspace{-2.2em}
        \subcaption{Setonix}
        \label{fig: Setonix speedup GEMM} 
     \end{subfigure}
    \vspace{-1.5em}
     \begin{subfigure}[b]{1\columnwidth}
         \centering
        \makebox[\columnwidth][c]{\includegraphics[width=1.06 \columnwidth]{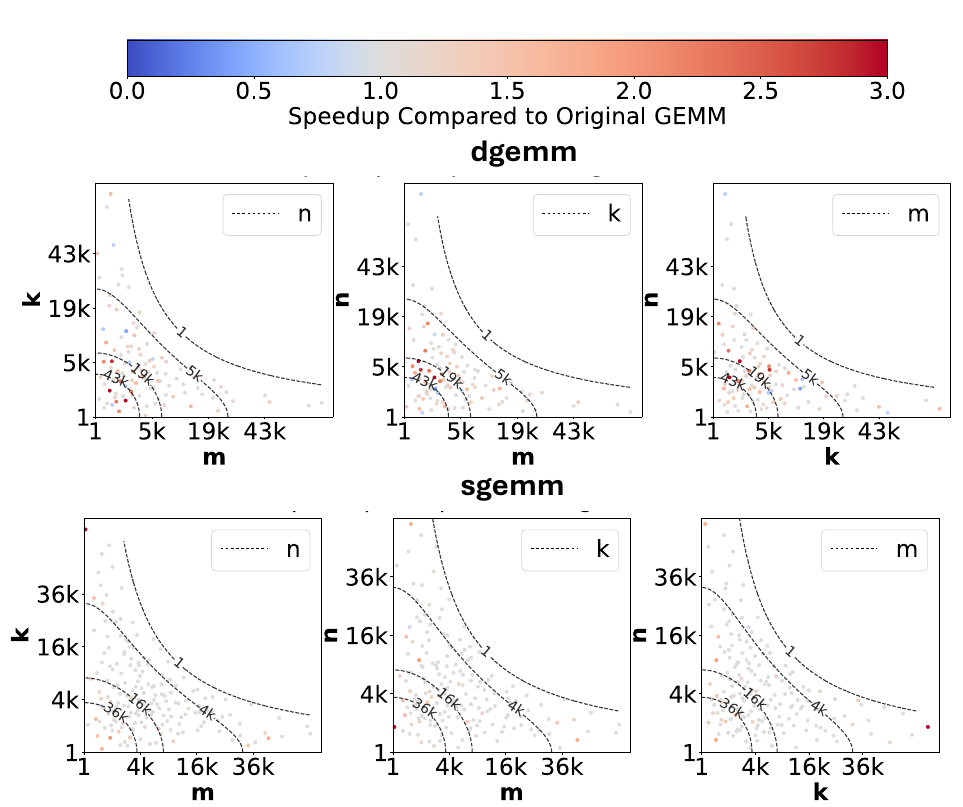}}
        \vspace{-2em}
        \subcaption{Gadi}
        \label{fig: Gadi speedup GEMM} 
        \vspace{1.5em}
     \end{subfigure}

  \caption{Heatmap for speedups with respect to matrix dimensions on Setonix and Gadi. The horizontal and vertical axes use a square root scale. The dashed lines on each sub-graph are contour lines of the sampling domain with each label showing the value of the third dimension.} 
  \label{fig: speedup GEMM} 
   \vspace{-1.0em}
\end{figure}


\begin{table}[!htbp]
  \centering
  \vspace{-0.5em}
  \caption{{Profiling results on Gadi with GEMM, SYMM, and SYRK.}}
  \label{tab: profiling Gadi}
  \scriptsize
\sffamily

\begin{tabular}{SSSSSSS}
    \toprule
     {$\textit{\textbf{m, k, n}}$} & \textbf{Thread Count} & \textbf{Total Time (s)} & \textbf{Thread Sync (s)} & \textbf{Kernel Call (s)} & \textbf{Data Copy (s)} \\

    \midrule

    {dgemm \textbf{${\text{64,2048,64}}$ no ML}}  & 96  &  11.001  &  4.806 &  0.010  &      2.060            \\
    {dgemm \textbf{${\text{64,2048,64}}$ with ML}}  & 16  &  4.430  & 1.453 &  0.001   &      0.887            \\

    \midrule

    {sgemm \textbf{${\text{64,2048,64}}$ no ML}}  & 96    &  7.698   &  2.716  &  0.005   & 1.133\\
    {sgemm \textbf{${\text{64,2048,64}}$ with ML}}  & 5    &   2.701    &  0.155  &  0.001   & 0.301\\

    \midrule
    \midrule
    {dsymm \textbf{${\text{248,39944}}$ no ML}}  & 96  &  29.063  &  11.627 &  0.496  &      9.783            \\
    {dsymm \textbf{${\text{248,39944}}$ with ML}}  & 25  &  22.067  & 4.962 &  0.441   &      9.698            \\

    \midrule

    {ssymm \textbf{${\text{2759,41681}}$ no ML}}  & 96  &  27.299  &  8.986 &  1.208  &      10.605            \\
    {ssymm \textbf{${\text{2759,41681}}$ with ML}}  & 12  &  15.597  & 2.085 &  0.241   &      8.254            \\

    \midrule
    \midrule

    {dsyrk \textbf{${\text{124,160163}}$ no ML}}  & 96  &  36.060  &  8.545 &  0.005  &      13.672            \\
    {dsyrk \textbf{${\text{124,160163}}$ with ML}}  & 43  &  34.371  & 7.543 &  0.001   &      12.529            \\

    \midrule

    {ssyrk \textbf{${\text{175,15095}}$ no ML}}  & 96  &  65.844  &  62.666 &  1.639  &      1.103            \\
    {ssyrk \textbf{${\text{175,15095}}$ with ML}}  & 46  &  4.581  & 1.323 &  1.744   &      1.263            \\

    \bottomrule
    \bottomrule
\end{tabular}






  \vspace{-1em}
\end{table}

\section{Performance Analysis}
\label{sec: Performance Analysis}
In this section, we first provide a visual representation and succinct explanation of our datasets. We then present the results of our machine learning model selection process. Finally, we evaluate the performance of our method and its software implementation using our test datasets. All analysis are performed on both Setonix and Gadi supercomputers and for all BLAS L3 subroutines with double and single precision.


\subsection{Datasets{, Training Time}, and Data Visualization}
\label{subsec: Data Collection}



The size of the training dataset is 1000-1200 for each of the BLAS III subroutines on each HPC platform. It is believed that the size of the dataset is sufficient, since it is generally observed that the validation performance does not improve significantly with more training set data.


The data gathering on Setonix and Gadi required approximately 100 node hours for each subroutine. It was observed that, given the same upper limit of matrix size (specifically 500 MB), the double precision subroutines took more time to collect data compared to their single precision counterparts.

Figure \ref{fig: data collection} contains two subfigures that depict the distribution of the optimal number of threads for GEMM on Setonix and Gadi, respectively. The patterns for sGEMM and dGEMM appear to be quite similar on Setonix, but show more variation on Gadi. A common trend observed is that irregular GEMM calls, where at least one of the dimensions $m$, $k$, or $n$ is small, often lead to suboptimal performance. Also, we can observe patches of abnormal area where choices of the optimal number of threads is drastically different from the surrounding area. 

The optimal number of threads for SYMM, SYRK, SYR2K, TRMM, and TRSM on Setonix and Gadi is displayed in Fig. \ref{fig: data collection 2d}. The patterns for double precision and single precision subroutines are mostly similar, but there are some exceptions. Also, the distribution of the optimal number of threads is more uneven than for GEMM. For SYRK, TRMM, and TRSM on Setonix, many calls have optimal number of threads higher than the number of physical cores (hyperthreads). On the contrary, for SYRK, SYR2K, and TRMM on Gadi, almost all calls have optimal number of threads lower than the number of physical cores. There are also abnormal areas where the optimal number of threads is drastically different from the surrounding area, similar to GEMM.

The complex performance patterns of BLAS level III subroutines are shown by these results; our ML models aim to learn these patterns and select the best number of threads.

As mentioned in Section \ref{subsec: Data Preprocessing}, we use stratified sampling to split the data set for model training and testing, with 15 $\%$ of the data set as the test set.

\subsection{Model Performance and Selection}

The best algorithms chosen by our algorithm of selecting the ML model with the highest average estimated speedup s for all BLAS subroutines are shown in Table \ref{tab:estimated_speedup_setonix} and \ref{tab:estimated_speedup_gadi}. We can see that only four algorithms are chosen at least once, with XGBoost still being the most common option. The results also indicate that the best algorithms for Setonix and Gadi are sometimes different, which is expected since the two platforms have different architectures. {We have also included in Table} \ref{tab:detailed stats speedup} {detailed statistics supporting the selection of most suitable models for four subroutines on Gadi.}

We regret that we have not completed integrating all four models in our code base. For now, we only use XGBoost as the ML model for both Setonix and Gadi. Since the estimated speedup only differ by less than 10 \%, we expect the speedup difference will not be significant. We will integrate the other three models in the future.

We used separate datasets for each subroutines from the training and test datasets test ADSALA on BLAS L3 subroutines. These datasets consist of 100-120 data points sampled with a scrambled Halton sequence within the same domain as how its training dataset is sampled. This is expected to ensure a more uniform, low discrepancy sampling of the data used for the performance analysis of our software. We compared the subroutine runtime with our ADSALA software to the runtime of original using maximum number of threads. Please note that the speedup results include the model evaluation time during runtime.

The testing shows that our ADSALA software can reliably improve the performance of BLAS L3 subroutines on both Setonix and Gadi. Table \ref{tab: performance} shows the statistics of the speedup results. The mean speedup on Setonix is generally higher with also higher standard deviation than on Gadi. Precision-wise, double precision subroutines generally have higher speedups than their single precision counterparts. Subroutine-wise, SYMM has the highest mean speedup on both platforms, while GEMM and SYR2K has the lowest mean speedup on Setonix and Gadi, respectively.





Figure \ref{fig: speedup GEMM} visualises the GEMM speedup distribution on Setonix and Gadi as a 3D heatmap from three angles, where red indicates acceleration and blue indicates deceleration. We can see that the speedup distribution pattern resembles the pattern of room for improvements, and the speedup generally decreases as three dimensions get larger.

Figure \ref{fig: 2dim speedup} shows the SYMM, SYRK, SYR2K, TRMM, and TRSM speedup distribution with respect to matrix dimensions on Setonix and Gadi. We can observe that the speedup patterns also resembles that in Fig. \ref{fig: data collection 2d}. The double/single subroutine pairs generally show similar speedup patter. However, speedup shows various patterns for different subroutines. Also, it does not always follows the rule that the speedup decreases as the matrix dimensions get larger.

\subsection{Performance Assessment using Software Implementation}

We use Table \ref{tab: profiling Gadi} to explain the large speedups of ADSALA over the original BLAS package in some cases. These cases have considerable speedup and are selected from the test set for software testing.

We use Intel$^{\circledR}AdvisorandIntel^{\circledR}$ Vtune to profile these two GEMMs on Gadi, repeating each matrix multiplication 100 times with different input values.
Table \ref{tab: profiling Gadi} shows the time breakdown of GEMM calls. The SGEMM wall-time consists of three main components:
\begin{enumerate}
\item Data copies. The BLAS runtime uses a buffer for each thread as a workspace and copies blocks of matrix operands into it before computation. The copy time depends on the matrix sizes, memory placement, and thread number.
\item Thread synchronization.
\item BLAS L3 kernel calls, where the FLOPs are performed. The kernels depend on the thread number and the matrix block sizes. They can be compute-bound for large blocks.
\end{enumerate}

Generally speaking, the speedup of ADSALA is due to the reduction of all three parts. Because of the relatively small problem size, the most significant time consumption is on thread synchronization most of the time. The thread sync time can be improved from around 30 \% to more than 50 times. Data copy has the second largest time consumption, and the speedup on it also contributes to the reduction of total runtime. Although the kernal call is consumes minimal time, our number threads selected by ML can also reduce the time consumption on it.



\section{Conclusions}
\label{sec: Conclusion}



We presented an extension to the ADSALA library where only GEMM was optimized. We extend its approach to fit other BLAS L3 routines with different input matrices. While keeping the same method for selecting the ML algorithm, we found the optimal algorithm is architecture and subroutine dependent.

Out of the six BLAS L3 subroutines, we improved on all of them, with speedups ranging from 1.1 to 3.0. We also analyzed visualise and analyse the speedup patterns of different BLAS subroutines and HPC platforms. By profiling, we explained the large speedups in some cases and showed that the source of speedup is the reduction of all three main components of the multi-thread subroutine runtime. 

{Our method is applicable to runtime parameter predictions, including multi-thread BLAS I, II, and LAPACK operations, which are sensitive to prediction duration. Future work will investigate automatic processor selection, such as the feasibility of running BLAS/LAPACK operations on either a GPU or CPU. At present, our approach is applicable to problems that satisfy certain conditions: a well-defined objective function, a finite and discrete search space, a set of relevant features, and sufficient data for training and validating ML models.}

Our work shows the effectiveness and generality of the ADSALA approach for optimizing BLAS routines on modern multi-core systems. For future works, we plan to extend our ML-driven runtime thread selection approach to other BLAS operations and to a more diverse set of computing hardware, including GPU accelerators and heterogeneous systems.





\bibliographystyle{IEEEtran}
\bibliography{thesis.bib}

\end{document}